\documentclass[useAMS,usenatbib]{mn2e}
\usepackage{graphicx}

\title[Automatic morphological classification of galaxy images]{Automatic morphological classification of galaxy images}

\author[Lior Shamir]{Lior Shamir$^{1}$\thanks{E-mail: shamirl@mail.nih.gov} \\
$^{1}$Laboratory of Genetics, NIA/NIH, 251 Bayview Boulevard, Baltimore, MD 21224, USA}

\begin{document}

\date{}

\pagerange{\pageref{firstpage}--\pageref{lastpage}} \pubyear{2005}

\maketitle

\label{firstpage}

\begin{abstract}

We describe an image analysis supervised learning algorithm that can automatically classify galaxy images. The algorithm is first trained using a manually classified images of elliptical, spiral, and edge-on galaxies. A large set of image features is extracted from each image, and the most informative features are selected using Fisher scores. Test images can then be classified using a simple Weighted Nearest Neighbor rule such that the Fisher scores are used as the feature weights. Experimental results show that galaxy images from {\it Galaxy Zoo} can be classified automatically to spiral, elliptical and edge-on galaxies with accuracy of $\sim$90\% compared to classifications carried out by the author. Full compilable source code of the algorithm is available for free download, and its general-purpose nature makes it suitable for other uses that involve automatic image analysis of celestial objects.
\end{abstract}


\maketitle

\section{Introduction}
In the past several years autonomous sky surveys have been becoming increasingly important, and large datasets of astronomical images have been generated and become available by these ventures. The availability of these large datasets has introduced the need for tools that can automatically analyze astronomical images. This includes the need for automatic morphological classification of celestial objects that appear inside an astronomical frame.

One approach to classification of large sets of galaxy images, which was successfully adopted by the {\it Galaxy Zoo} project \citep{Lin08}, allows hobbyist volunteers to log-in and manually classify galaxies via the project web site. The galaxy images are acquired by the Sloan Digital Sky Survey (SDSS), and displayed by {\it Galaxy Zoo} as JPEG images scaled by 0.024$R_p$, where $R_p$ is the Petrosian radius \citep{Pet76} for the galaxy. 

While each volunteer can classify just a limited number of galaxies, the efficacy of the data analysis is enabled by the availability of a very large number of human observers. However, the bottleneck introduced by the manual analysis limits the ability of this method to provide quick analysis of massive galaxy datasets.

Here we describe a software tool that can be used for automatic classification of galaxy images. The algorithm was originally developed for automatic analysis of cell morphology, but its general-purpose design allows it to be effective for applications outside the scope of cell biology. Full compilable source code can be freely downloaded. In Section~\ref{method} we briefly describe the algorithm, and in Section~\ref{results} the experimental results are discussed.

\section{Image analysis method}
\label{method}

The image analysis algorithm used for the automatic galaxy image classification is {\it WND-CHARM} \citep{Sha08a,Orl08}, which was originally designed for automatic analysis of cell and tissue images, but also demonstrated efficacy as a general-purpose image analysis tool \citep{Sha08,Sha09a}. {\it WND-CHARM} first reduces each image to a total of 2873 numerical low-level descriptors (when the ``-l" switch in the command line is turned on, which indicates that the larger set of image features should be computed). These generic image features include high-contrast features (object statistics, edge statistics, Gabor filters), textures (Haralick, Tamura), statistical distribution of the pixel values (multi-scale histogram, first four moments), factors from polynomial decomposition of the image (Chebyshev statistics, Chebyshev-Fourier statistics, Zernike polynomials), Radon features and fractal features. A detailed description of these image content descriptors is available in \citep{Orl08,Sha08,Sha08a,Sha08b,Sha09a,Sha09b}.  To extend the number and variety of the image features, these algorithms are applied not only to the raw image pixels, but also to several transforms of the image such as Fourier, Chebyshev, Wavelet, and edge-magnitude transform, as well as tandem transform combinations \citep{Sha08a,Sha08}.  

After image content descriptors for all images in the training dataset are computed, each of the 2873 features is assigned a Fisher discriminant score \citep{Bis06}, and 85\% of the features with the lowest Fisher scores are rejected in order to filter non-informative image features. The distance between two image feature vectors {\it X} and {\it Y} can then be computed by using a simple Weighted Nearest Neighbor rule, as described by Equation~\ref{wnn}

\begin{equation}
d=\sqrt{\sum_{f=1}^{|X|} W_f(X_f-Y_f)^2},
\label{wnn}
\end{equation}
where $W_f$ is the assigned Fisher score of feature {\it f}, and {\it d} is the computed weighted distance between the two feature vectors. The predicted class of a given test image is simply determined by the class of the training image that has the shortest weighted distance {\it d} to the test image. A compilable open-source of the {\it WND-CHARM} algorithm is available for free download at http://www.cs.mtu.edu/$\sim$lshamir/downloads/ImageClassifier.

\section{Experimental results}
\label{results}

The method was tested using a dataset of spiral and elliptical galaxy images taken from {\it Galaxy Zoo} and classified manually by the author. The galaxies in the dataset were selected randomly by the {\it Galaxy Zoo} web interface, and no attempt to normalize for luminosity, size or distance was made. Unclear cases were classified by the judgment of the author. This study, however, ignored {\it Galaxy Zoo} monochrome images, that were introduced by the {\it Galaxy Zoo} bias study \citep{Lin08}. Although only colour images were used, no colour features were used in this study.

The 120$\times$120 pixel block at the centre of each galaxy image was separated from the image and converted into lossless TIFF image format, from which image content descriptors were computed. The dataset includes images of 247 spiral galaxies (one is repeated), 215 elliptical galaxies, and 107 edge-on galaxies, and can be downloaded at http://www.cs.mtu.edu/$\sim$lshamir/downloads/galaxies.tar.gz.

In the first experiment, the image classification method was used to classify between spiral and elliptical galaxies. One hundred and fifty images from each class (spiral and elliptical) were used for training, and 50 images for testing (by specifying the ``-i150" and ``-j50" parameters in the {\it wndchrm} command line). The experiment was repeated 30 times such that in each run different images were selected randomly from the pool of images and were allocated for the training and test sets. The results show that $\sim$93\% of the galaxy images were classified correctly to elliptical and spiral galaxies, as can be learned from the confusion matrix of Table~\ref{confusion_matrix}.

\begin{table}
 \centering
 \begin{minipage}{70mm}
  \caption{Confusion matrix of the classification of elliptical and spiral galaxies}
  \label{confusion_matrix}
  \begin{tabular}{@{}llc|c@{}}
  \hline
  &                   Elliptical & Spiral \\
  \hline
  Elliptical &  1445         &  55     \\
  Spiral      &   150          & 1350 \\
 \hline
\end{tabular}
\end{minipage}
\end{table}

\begin{figure*}
\includegraphics[angle=90]{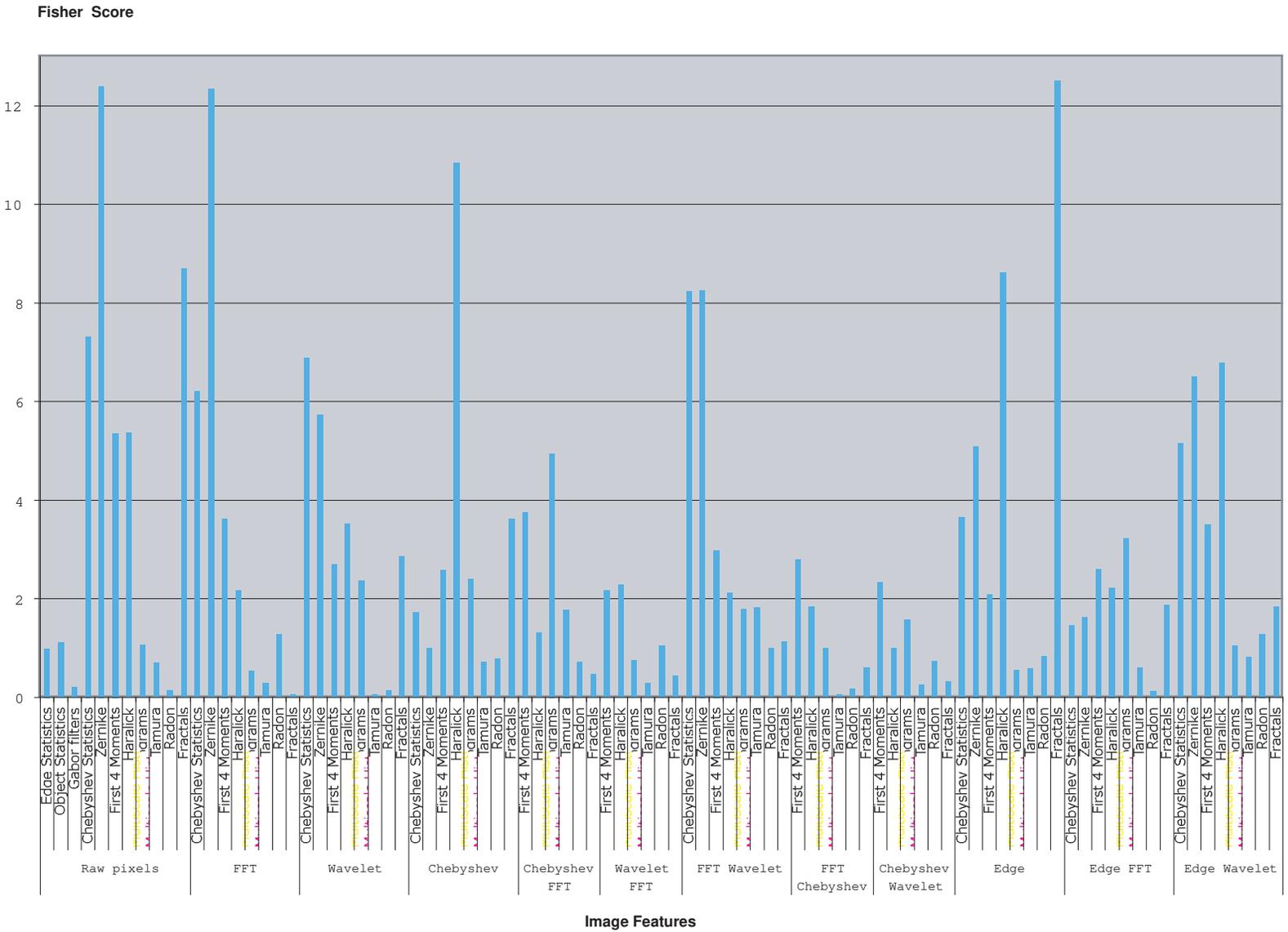}
\caption{Fisher scores of the image features computed on the different image transforms and compound image transforms}
\label{features}
\end{figure*}

While {\it WND-CHARM} computes a large set of image features, not all features are expected to be equally informative, and some are expected to represent noise. The estimated informativeness of the different image content descriptors is described by Figure~\ref{features}, which shows the sum of the Fisher scores of all bins of the different feature groups extracted from the different image transforms \citep{Sha08a}.

While some of the informative image features are highly non-intuitive, such as the Haralick texture features \citep{Har73} computed from the Chebyshev image transform, other image content descriptors are easier to conceptualize. For instance, the fractal features used by {\it WND-CHARM} \citep{Wu92} can become informative by sensing the fractal characteristics of the shape of a spiral galaxy, which are not expected to exist in an elliptical galaxy. The fractality of spiral galaxies can often be sensed easily by the unaided eye. One example is the picture of the M101 ``pinwheel" galaxy \citep{Nem09}, in which some of the arms split into secondary arms, which then split again to smaller arms. When using the fractal features alone, the classification accuracy between the spiral and elliptical galaxies is $\sim$76\%, which demonstrates the informativeness of the fractal features for galaxy morphology.

Clearly, the M101 picture taken by Hubble Space Telescope is much more detailed than the galaxy images acquired by SDSS. However, while the fractality signal is obviously weaker in the Sloan images, it still exists. Fractal analysis methods can very often detect fractality that is very difficult to sense by the unaided eye, and is sensitive to even subtle fractal patterns \citep{Man82}. Therefore, the informativeness of the fractal features for detecting spiral galaxies in the small-scale SDSS images cannot be considered surprising.

Other informative features include the Zernike polynomials \citep{Tea79}, which are also expected to be informative due to their radial nature that allows them to reflect variations in the unit disk. Since the unit disk is definitely a fundamental and obvious difference between spiral and elliptical galaxies, Zernike polynomials are expected to reflect differences between these types of galaxies. Zernike polynomial features can also be used for classification between true elliptical galaxies and S0 galaxies that have a disk, which are a major source of confusion in morphological classification of galaxies. When only the Zernike features are used for the classification, the accuracy is $\sim$71\%.

In addition to the predicted class of the galaxy ({\it spiral} or {\it elliptical}), the classifier also returns the similarity of the tested galaxy images to each of the classes, measured by the distances between the feature vectors as described in Section~\ref{method}, normalized to the interval $(0,1)$. For instance, for a galaxy that is clearly spiral the similarity values to the classes {\it spiral} and {\it elliptical} are expected to be relatively different from each other such as 0.65 and 0.35, respectively. For a galaxy that does not have an obvious typical spiral shape, the two values are expected to be more similar, such as 0.52 and 0.48. These similarity values can provide additional information about the morphology of the galaxies which aims to measure {\it how} spiral or elliptical they are. Table~\ref{galaxy_images} shows some sample galaxy images with their computed similarity values and the automatic and manual classifications, including some cases of disagreement between the author's and the automatic classification. While the estimated similarity for a single image is not always accurate, large sets of images of each type can allow quantitative analysis of the similarity between the different classes \citep{Sha08a,Sha09a}.

\begin{table*}
 \centering
 \begin{minipage}{140mm}
  \caption{Automatic and author classification of sample {\it Galaxy Zoo} galaxy images}
  \label{galaxy_images}
  \begin{tabular}{@{}llc|c|c|c@{}}
  \hline
     Image & Galaxy Zoo ID & Similarity values & Automatic  & Author's  \\
                 &                            &  (elliptical/spiral) & classification &  classification \\
  \hline
  
\includegraphics[scale=0.5]{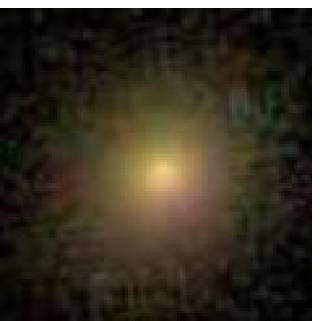} & 588023669702131872 &  0.547/0.453 & elliptical & elliptical \\
\includegraphics[scale=0.5]{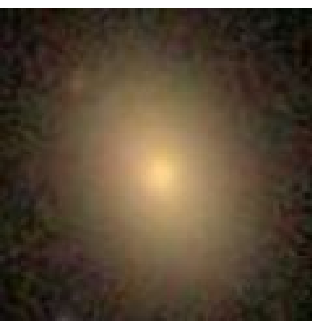} & 587739380994998479 &  0.562/0.438 & elliptical & elliptical \\
\includegraphics[scale=0.5]{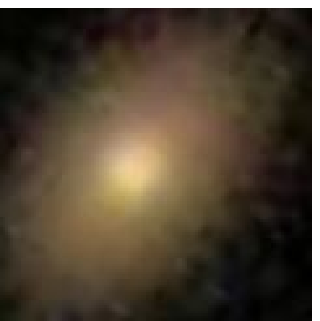} & 587736919969890614 &  0.520/0.480 & elliptical & elliptical \\
\includegraphics[scale=0.5]{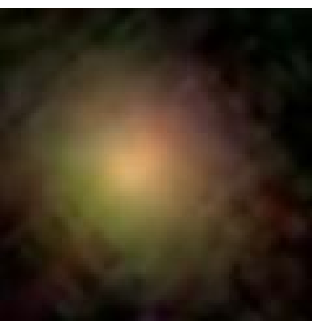} & 587742783673991349 &  0.513/0.487 & elliptical & elliptical \\
\includegraphics[scale=0.5]{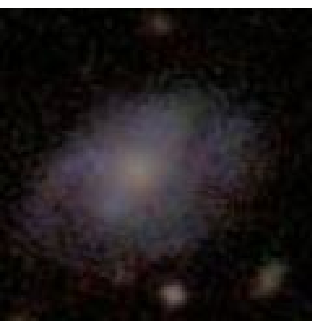} & 587742575925657806 &  0.466/0.534 & spiral & spiral \\
\includegraphics[scale=0.5]{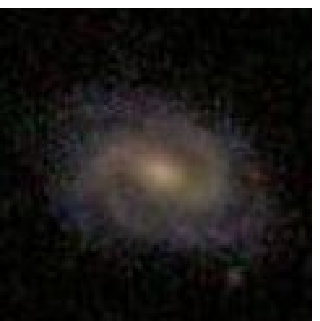} & 588017721180881084 &  0.436/0.564 & spiral & spiral \\
\includegraphics[scale=0.5]{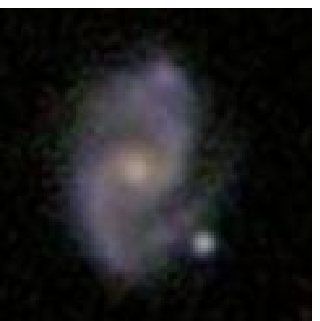} & 587736585508094159 &  0.416/0.584 & spiral & spiral \\
\includegraphics[scale=0.5]{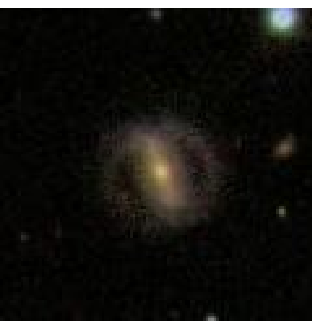} & 588009366939238541 &  0.488/0.512 & spiral & spiral \\
\includegraphics[scale=0.5]{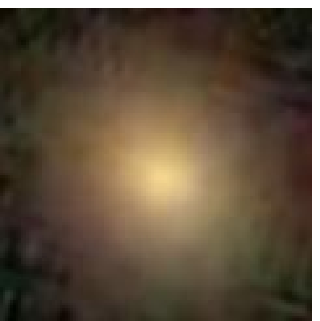} & 587735697522229435 &  0.508/0.492 & elliptical & elliptical \\
\includegraphics[scale=0.5]{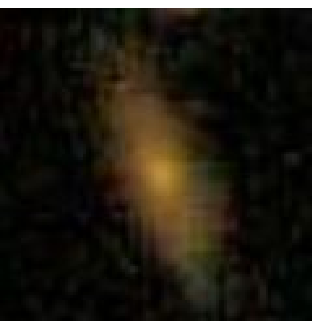} & 587727221950447853 &  0.504/0.496 & elliptical & spiral \\
\includegraphics[scale=0.5]{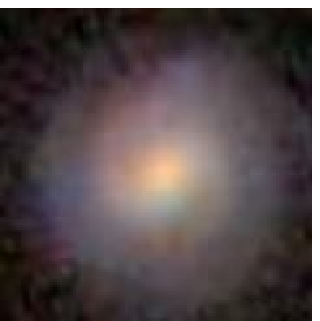} & 587741600950452406 &  0.510/0.490 & elliptical & spiral \\
\includegraphics[scale=0.5]{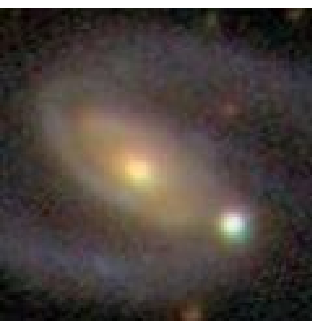} & 588016878292762809 &  0.502/0.498 & elliptical & spiral \\
\includegraphics[scale=0.5]{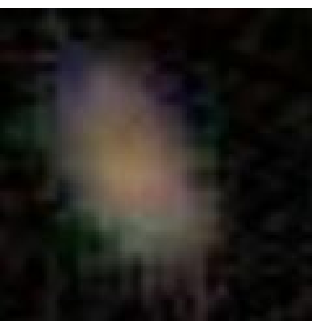} & 588015509808152736 &  0.427/0.573 & spiral & elliptical \\

\hline
\end{tabular}
\end{minipage}
\end{table*}

\begin{figure}
\includegraphics[angle=0,scale=0.43]{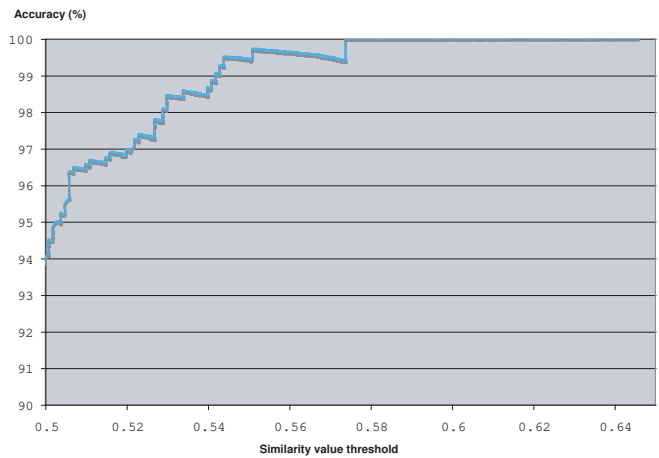}
\caption{Classification accuracy when excluding galaxy images with similarity lower than a certain threshold}
\label{sim_threshold}
\end{figure}

The similarity value can also be used as an indication of the certainty of a galaxy image classification. i.e., a classification of a galaxy image with a high similarity value to a certain morpohlogical type can be considered more certain than a classification in which the similarity value is slightly greater than 0.5. Figure~\ref{sim_threshold} shows how the classification accuracy responds to threshold similarity values. As the figure shows, all galaxy classifications with similarity values greater than 0.58 were classified correctly, and the classification accuracy is $\sim$98.5\% for galaxy image classifications with similarity values greater than 0.54.

\begin{figure}
\includegraphics[angle=0,scale=0.44]{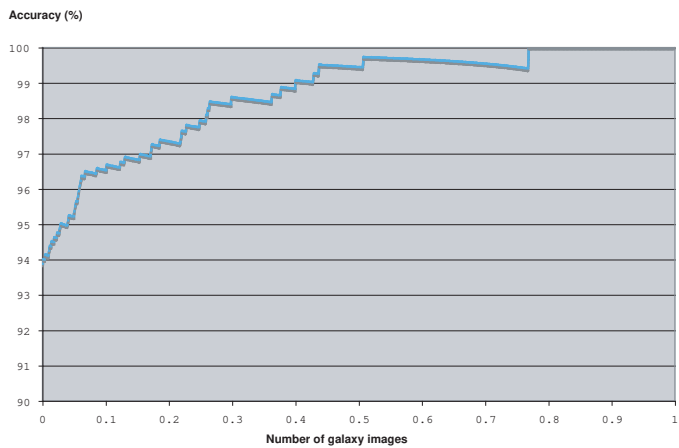}
\caption{Classification accuracy as a function of the number of galaxies with the highest similarity values}
\label{amount_threshold}
\end{figure}

The amount of galaxy images that correspond to the threshold similarity values is shown in Figure~\ref{amount_threshold}.  As the figure shows, $\sim$50\% of all galaxy images can be classified with accuracy greater than 99.5\%, and $\sim$80\% of the galaxies with accuracy greater than 97\%. When counting also the galaxy image classifications that have similarity values slightly higher than 0.5 the classification accuracy drops below 94\%.

An additional experiment tested whether the image classifier described in Section~\ref{method} can classify between edge-on and spiral galaxies. For this purpose, 80 images of edge-on galaxies and a similar number of spiral galaxies were used for training, and 20 images from each class for testing (by using the ``-i80" and ``-j20" parameters of the {\it wndchrm} command line). The experiment was repeated 50 times, such that in each run images were allocated randomly to training and test sets. The results show that 92\% of the images were classified correctly to spiral and edge-on galaxies, as can be learned from the confusion matrix of Table~\ref{confusion_matrix2}.

\begin{table}
 \centering
 \begin{minipage}{70mm}
  \caption{Confusion matrix of the classification of edge-on and spiral galaxies}
  \label{confusion_matrix2}
  \begin{tabular}{@{}llc|c@{}}
  \hline
  &                   Edge-on & Spiral \\
  \hline
  Edge-on &  929         &  71     \\
  Spiral      &   85          & 915    \\
 \hline
\end{tabular}
\end{minipage}
\end{table}

A similar experiment tested whether the image classifier can classify between elliptical and edge-on galaxies. In this experiment, the dataset included 100 images of elliptical galaxies and 100 images of edge-on galaxies, such that 80 images from each set were used for training and 20 for testing. As before, the experiment was repeated 50 times, and the average classification accuracy was 98\% as shown by the confusion matrix of Table~\ref{confusion_matrix3}.

\begin{table}
 \centering
 \begin{minipage}{70mm}
  \caption{Confusion matrix of the classification of edge-on and elliptical galaxies}
  \label{confusion_matrix3}
  \begin{tabular}{@{}llc|c@{}}
  \hline
  &                   Edge on & Elliptical \\
  \hline
  Edge on      &  976         &  24     \\
  Elliptical      &   8          & 992    \\
 \hline
\end{tabular}
\end{minipage}
\end{table}

The accuracy of a three-way classifier for all three classes together (spiral, elliptical and edge-on galaxies) is 90\%. This was determined by 30 runs, such that in each run 80 images of each of the three classes were randomly selected for training, and 20 images for testing. The confusion matrix of the experiment is described by Table~\ref{confusion_matrix4}.

\begin{table}
 \centering
 \begin{minipage}{70mm}
  \caption{Confusion matrix of the classification of edge-on, elliptical, and spiral galaxies}
  \label{confusion_matrix4}
  \begin{tabular}{@{}llc|c|c@{}}
  \hline
  &                   Edge-on & Elliptical & Spiral \\
  \hline
  Edge-on &  551     &  7        &   42 \\
  Elliptical  &     5      &  561   &    34 \\
  Spiral      &   38      & 47       &   515 \\
 \hline
\end{tabular}
\end{minipage}
\end{table}

\begin{figure}
\includegraphics[angle=0,scale=0.50]{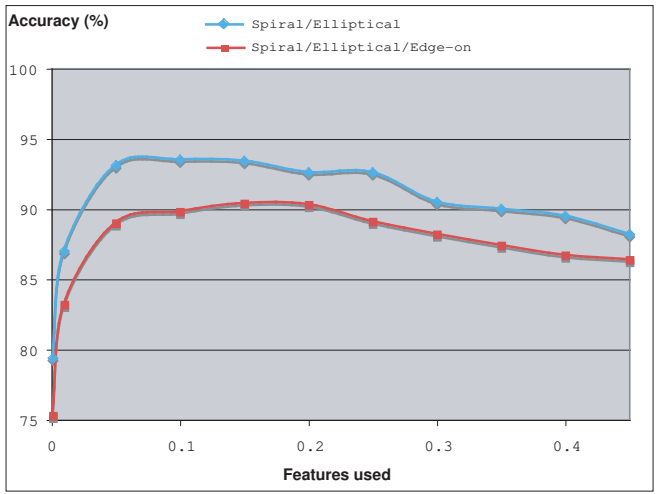}
\caption{Classification accuracy as a function of the size of the feature set}
\label{fig2}
\end{figure}

As discussed in \citep{Orl08,Sha09a}, the accuracy of the image classifier is not very sensitive to the number of the image features due to the use of the feature weights when computing the distances between feature vectors. Rejecting the weakest 85\% of the features when using the larger feature set of the {\it wndchrm} tool \citep{Sha08a} is often a reasonable starting point, and in many cases other values (set by using the ``-f" option in the command line) do not improve the performance significantly. Figure~\ref{fig2} shows how the classification accuracy changes as more features are used.

As the figure shows, the classification accuracy increases as the number of used features gets larger, and starts to decrease when more than 15\% of the features are used due to the increasing effect of noisy features.

A major downside of the proposed algorithm is its computational complexity. The extraction of a large number of 431 image features (15\% of 2873) from each image is a computationally intensive task, so that computing the image features for a single image takes $\sim$35 seconds using a system with a 2GHZ Intel Processor and 2GB of RAM. However, the step of image feature extraction can be parallelized with a very low overhead \citep{Sha08a}, so that several processors can compute the same dataset, reducing the response time of the system almost linearly to the number of processors. The classification of the feature values cannot be parallelized without changing the software, but the computational cost of this step is negligible.

The performance of the proposed image analysis method was compared to galaxy classification using the Gini coefficient \citep{Abr03}. This was done by using the {\it morph} command-line utility, which is part of the {\it Morpheus} package \citep{Abr03}. The same set of galaxy images was used, but the images were converted into FITS format, which is the native input format of {\it morph}. Results show that in $\sim$77\% of the cases the Gini coefficient accurately determined whether a galaxy is elliptical or spiral, and Table~\ref{confusion_matrix_gini} shows the confusion matrix of the classification. The agreements between the method proposed in this paper and the Gini coefficient method is $\sim$75\% for the elliptical galaxies, and $\sim$66\% for the spiral galaxies. Clearly, there is a better degree of agreement between the two methods on elliptical galaxies comparing to spiral galaxies. When using the Gini coefficient method for classifying between the three types of galaxies (elliptical, spiral and edge-on), the classification accuracy is $\sim$55\%, as can be learned from the confusion matrix of Table~\ref{comfusion_matrix_gini_3way}.

It should be noted that the computational cost of the Gini coefficient, which is practically negligible, makes it dramatically faster than computing the set of image features used in this paper. It should also be noted that the Gini coefficient performed better than any other single image content descriptor included in the tested feature set \citep{Sha08a}. 

\begin{table}
 \centering
 \begin{minipage}{70mm}
  \caption{Confusion matrix of classification of elliptical and spiral galaxies using the Gini coefficient}
  \label{confusion_matrix_gini}
  \begin{tabular}{@{}llc|c@{}}
 \hline
  &                   Elliptical & Spiral \\
  \hline
  Elliptical &  157        &  43        \\
  Spiral      &   51         & 149       \\
 \hline
\end{tabular}
\end{minipage}
\end{table}

\begin{table}
 \centering
 \begin{minipage}{70mm}
  \caption{Confusion matrix of the classification of edge-on, elliptical, and spiral galaxies using the Gini coefficient}
  \label{comfusion_matrix_gini_3way}
  \begin{tabular}{@{}llc|c@{}}  
 \hline
   &                   Edge-on & Elliptical & Spiral \\
  \hline
  Edge-on  &  52    &   31    &   17 \\
  Elliptical  &   28      &   55  &   17  \\
  Spiral      &   22      &    20    &   58 \\
 \hline
\end{tabular}
\end{minipage}
\end{table}

\section{Conclusions}

Here we described an algorithm that can automatically classify between images of spiral, elliptical, and edge-on galaxies. The galaxy dataset features a random collection of galaxy images. Since luminosity, size and distance was found highly important for the automatic classification of galaxies \citep{Bam09}, it can be assumed that the classification accuracy can be improved when using datasets of nearby, large or bright galaxies. Since the described supervised machine learning method can be used for general purpose image classification, it is reasonable to assume that the same utility can be used for other problems in morphological analysis of celestial objects. 

The native format of Sloan images is FITS. Since the conversion from FITS format to lossy JPEG requires the sacrifice of image information, it can be assumed that direct access to the raw Sloan image files can potentially lead to a better performance, especially in cases of subtle differences of pixel intensity. Researchers are therefore advised to take this issue under consideration when applying {\it WND-CHARM} to problems in automatic galaxy morphology in which the differences between the galaxies are more difficult to notice by the unaided eye.

The dataset used for the described experiments consists of galaxy images manually classified by the author. Since supervised learning is used, the classifier can be biased by the intuition of the person(s) who prepare the gold standard training data. Therefore, training data for an image classifier that can be used for practical galaxy morphology classification should be selected and reviewed carefully. Even if the selection of the data follows a different intuition than the author's, as long as the classification criteria are consistent for all images the supervised learning is expected to provide performance figures that are comparable to the results reported in this paper.

Interestingly, many of the challenges of automatic morphology analysis of galaxies appear to be quite similar to automatic analysis of cell morphology. For instance, the interest in automatic detection of binucleate galaxies, which indicate that the two galaxies are being merged, coincides with the interest in automatic detection of binucleate cells, which means that the cell failed to complete the process of mitosis (e.g., G1 arrest). Another example is the interest in peculiar galaxies, which coincides with the interest in affected cells or unexpected phenotypes that are found among very many regular cells.

One of the major advantages of the algorithm is that its full source code is available for free download as a compilable software package \citep{Sha08a} that has been tested for robustness and correctness, and researchers who have basic computer skills can easily use the application as a command line utility. Therefore, in cases where there is a need for computer-based morphological analysis we encourage scientists to try {\it WND-CHARM} before taking the labour-intensive challenge of designing, developing and testing new task-specific image classifiers.

Applications of this method to galaxy classification include fully automatic analysis of galaxies, but it can also be used as a decision-supporting tool for datasets that are classified manually such as {\it Galaxy Zoo}.

\section{Acknowledgments}

This research was supported entirely by the Intramural Research Program of the NIH, National Institute on Aging. I would also like to thank Roberto Abraham for sharing the {\it Morpheus} code, and the referee, Chris Lintott, for his insightful comments.

Funding for the SDSS and SDSS-II has been provided by the Alfred P. Sloan Foundation, the Participating Institutions, the National Science Foundation, the US Department of Energy, the National Aeronautics and Space Administration, the Japanese Monbukagakusho, the Max Planck Society, and the Higher Education Funding Council for England. The SDSS Web Site is http://www.sdss.org/.

The SDSS is managed by the Astrophysical Research Consortium for the Participating Institutions. The Participating Institutions are the American Museum of Natural History, Astrophysical Institute Potsdam, University of Basel, University of Cambridge, Case Western Reserve University, University of Chicago, Drexel University, Fermilab, the Institute for Advanced Study, the Japan Participation Group, Johns Hopkins University, the Joint Institute for Nuclear Astrophysics, the Kavli Institute for Particle Astrophysics and Cosmology, the Korean Scientist Group, the Chinese Academy of Sciences (LAMOST), Los Alamos National Laboratory, the Max Planck Institute for Astronomy (MPIA), the Max Planck Institute for Astrophysics (MPA), New Mexico State University, Ohio State University, University of Pittsburgh, University of Portsmouth, Princeton University, the United States Naval Observatory and the University of Washington.

\label{last_page}

\end{document}